# A SECOND-GENERATION COSMIC AXION EXPERIMENT

C. Hagmann, W. Stoeffl, K. van Bibber
*Lawrence Livermore National Laboratory*
*7000 East Ave, Livermore, CA 94550*

E. Daw, D. Kinion, L. Rosenberg
*Dept of Physics, Massachusetts Institute of Technology*
*77 Massachusetts Ave, Cambridge, MA 02139*

P. Sikivie, N. Sullivan, D. Tanner
*Dept of Physics, University of Florida*
*Gainesville, FL 94720*

D. Moltz
*Nuclear Science Division, Lawrence Berkeley Laboratory*
*1 Cyclotron Rd, Berkeley, CA 94720*

F. Nezrick, M. Turner
*Fermi National Accelerator Laboratory*
*Batavia, IL 60510*

N. Golubev, L. Kravchuk
*Institute for Nuclear Research of the Russian Academy of Sciences*
*60th October Anniversary Prospekt 7a, Moscow 117312, Russia*

**Abstract**

An experiment is described to detect dark matter axions trapped in the halo of our galaxy. Galactic axions are converted into microwave photons via the Primakoff effect in a static background field provided by a superconducting magnet. The photons are collected in a high Q microwave cavity and detected by a low noise receiver. The axion mass range accessible by this experiment is 1.3 - 13 µeV. The expected sensitivity will be rougly 50 times greater than achieved by previous experiments in this mass range. The assembly of the detector is well under way at LLNL and data taking will start in mid-1995.

## Introduction

The axion is a hypothetical particle proposed to solve the strong CP problem in QCD. Peccei and Quinn [1] introduced a new U(1) symmetry into QCD which is broken at a scale $f_a$. The resulting Goldstone boson was called the axion [2]. Mass and couplings of the axion are inversely proportional to $f_a$. A priori, the energy scale $f_a$ is a free parameter of the theory, but a host of cosmological, astrophysical, and accelerator based constraints [3] limit the axion mass to 1 µeV< $m_a$ <1 meV. For 1 µeV< $m_a$ <100 µeV, axions may constitute some or all of the dark matter in the universe. At our location in the Milky Way the dark matter density is estimated to be 300 MeV/cm$^3$. Recent measurements [4] indicate that only about 20% of the halo material is baryonic, leaving the axion as a strong contender for the remainder of it.

In the past, two pilot experiments at Brookhaven [5] and Florida [6] have searched for galactic axions in the 4-16 µeV mass range, but lacked the sensitivity to detect axions with couplings as predicted by two plausible models, the KSVZ [7] and DFSZ [8] axion. The goal of the second-generation experiment under construction at LLNL is to find or exclude KSVZ axions over some mass range.

## Principle of Detection

In 1983, Sikivie proposed detecting galactic axions in a tunable microwave cavity permeated by a static magnetic field [9]. Galactic axions in a magnetic field have a small probability of decaying into single photons of frequency hf = $m_a c^2$ with a fractional spread of order 10$^{-6}$ due to kinetic energy. This conversion is resonantly enhanced by the quality Q of the cavity if the resonance frequency is tuned to the axion frequency (1 GHz =4.135 µeV). For DSFZ axions, the calculated power from axion to photon conversion into cavity mode $n$, and scaled to our experimental parameters, is given by

$$P_{a\rightarrow\gamma} = 9\times 10^{-23} \text{W} \left(\frac{B}{8\text{T}}\right)^2 \left(\frac{V}{200l}\right)\left(\frac{C_n}{0.7}\right)\left(\frac{Q_n}{10^5}\right)\left(\frac{m_a}{2\mu\text{eV}}\right)\left(\frac{\rho_a}{300\text{MeV}/\text{cm}^3}\right) \quad (1)$$

where $B$ is the static field, $V$ is the cavity volume, $\rho_a$ is the local axion density, and $C_n$ is a form factor related to the alignment of $B$ with the electric field of mode $n$. For KSVZ axions, P is about 7 times larger. To detect such small powers, the background noise must be sufficiently reduced. This is achieved by cooling the cavity to T ≈ 1.5 K and employing a cooled low noise amplifier with a noise temperature of $T_n$ ≈ 3 K. Axion signals are searched for by step tuning the cavity and looking for excess power. Figure 1 shows what a axion signal above background may look like. To achieve a reasonable search rate, the integration time at each frequency is limited to a few 10's of seconds. Initially, we are planning to scan about 500 MHz/yr with a signal to noise ratio s/n= 4 for KSVZ axions.

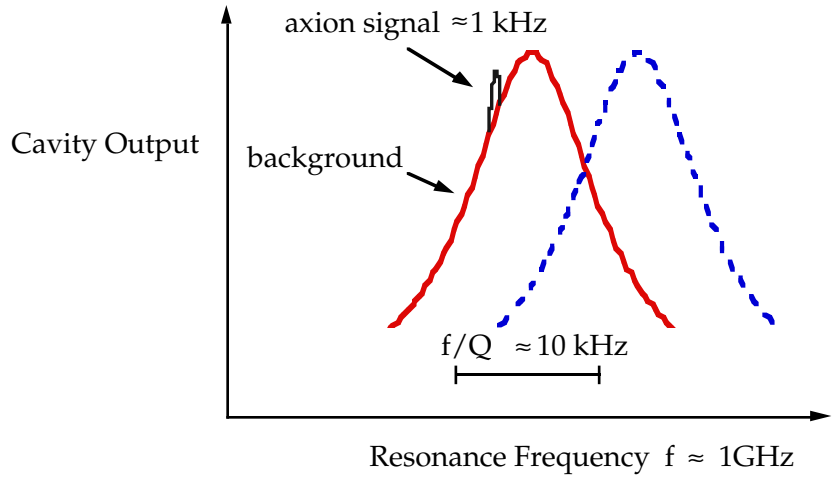

Fig.1: Axion signal and background. The dashed curve represents the cavity output after step tuning of the resonance frequency.

## Experimental Apparatus

A diagram of our experiment is shown in Figure 2. The outer dewar contains the NbTi magnet which has a central field of 8 T at a current of 235 A . The cavity and amplifier are housed in the inner dewar and cooled to T=1.5 K by pumping on LHe. The stainless steel cavity has a volume of 200 liters and was electroplated with 0.25 mm of copper and then annealed in UHV at 400°C for 8 hours to increase its low temperature RF conductivity. Tuning is accomplished by moving low-loss alumina rods ($\varepsilon \approx 10$) or hollow copper rods sideways in the cavity. The rods are driven by stepper motors on top of the cryostat. A double worm gear mounted on top of each rod reduces the steps by 1:30000 and allows low backlash frequency stepping. The total tuning range of the first cavity is 300-800 MHz (1.2-3.3 µeV) Figure 3 shows the tuning curve of the $TM_{010}$ mode with a single copper tuning rod. By changing tuning rods we will be able to cover the rest of the frequency range. For frequencies above 800 MHz, we plan to fill the available volume with arrays of 4 and 16 smaller cavities, with outputs combined in phase before amplification.

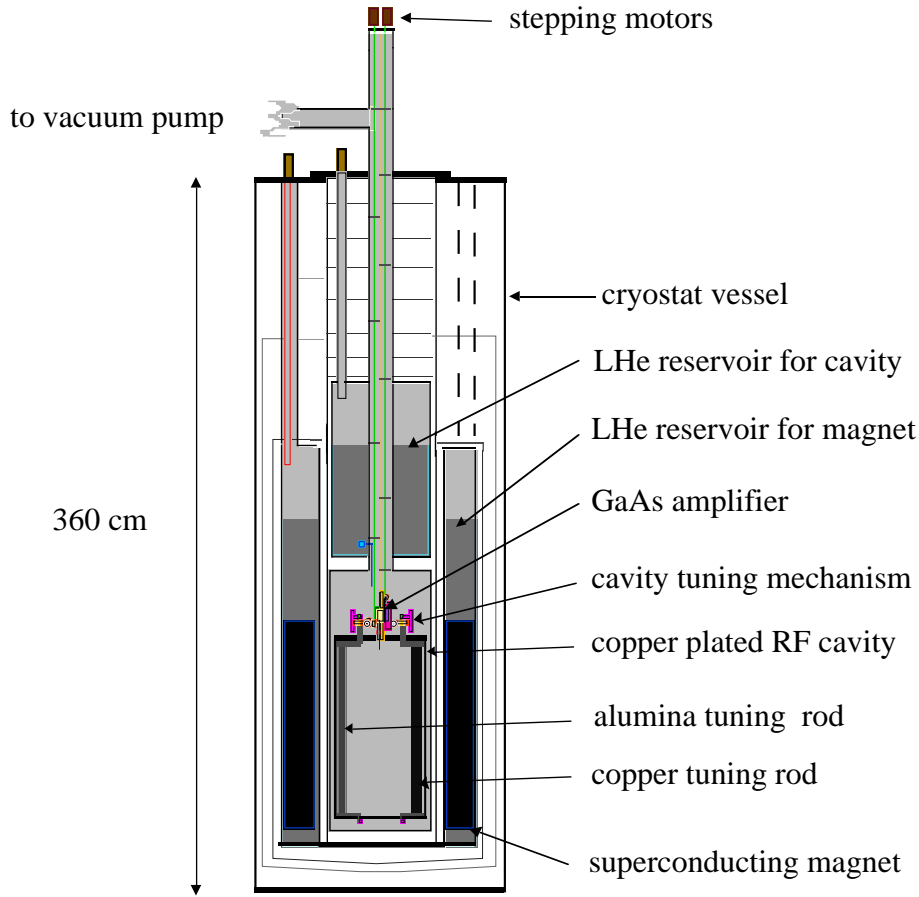

Fig.2: Layout of the axion experiment.

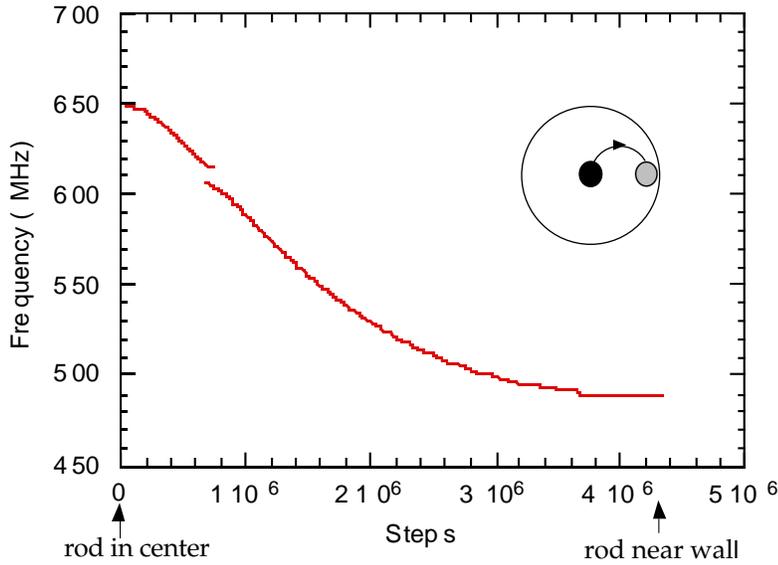

Fig.3: Tuning curve of the $TM_{010}$ mode with a single copper rod. The inset shows the cross section of the cavity.

## Receiver

In general, the noise temperature of a microwave receiver is determined by the front end amplifer. We use cooled two-stage GaAs amplifiers with $T_n \approx 3$ K below 1 GHz and 3-stage HEMTs with $T_n \approx 1$ K (f/1 GHz) above. Typically the bandwidth is 20 % and the power gain 35 dB. At room temperature, a double superheterodyne receiver mixes the cavity signal down to audio. A real time 400 point FFT analyzer calculates the power spectrum with a resolution of $\approx 50$ Hz. In parallel a high resolution ($\approx 0.1$ Hz) spectrum with about 1 million points will be taken by a PC based DSP board and searched for extremely narrow axion lines [10].

# Outlook

Our expected sensitivity by means of an exclusion plot is shown in Figure 4. In three years of running, we can exclude (or find) KSVZ axions provided our halo is saturated by axions. Still, to do a definitve search for DSFZ axions requires power sensitivities an order of magnitude better than at present. To this effect, we have started a collaboration with J. Clarke at UC Berkeley to develop DC SQUID amplifiers for microwave signals. Extrapolations of results obtained at 100 MHz and T = 1.5 K predict a noise temperature $T_n$ = 0.1 K at 500 MHz and T = 0.3 K. The system noise with the cavity cooled to T=0.3 K as well would then be $T_s = T_n + T_{phys}$ = 0.4 K. Design work has already begun on a dilution refrigerator and a compensation magnet for shielding the SQUID.

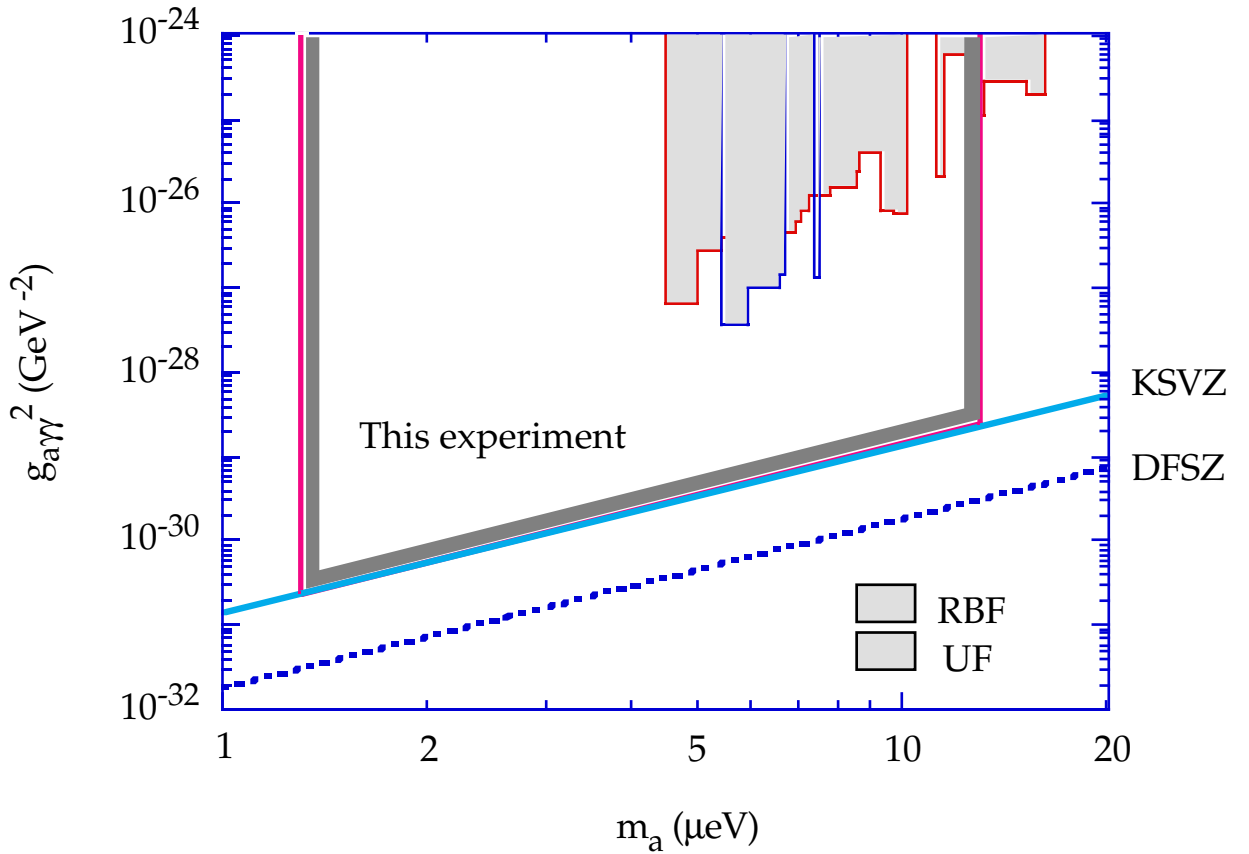

Fig.4: Exclusion plot of axion couplings and masses assuming that the halo is axionic. The hatched regions indicate the phase space excluded by the pilot experiments at Brookhaven and Florida.

# Acknowledgements

This work was performed under the auspices of the U.S. Department of Energy under contracts no. W-7405-ENG-48 (LLNL), DE-AC03-76SF00098 (LBL), DE-AC02-76CH03000